\magnification = \magstep1
\baselineskip = 20pt

\noindent{\bf THEORY OF THE SPATIO-TEMPORAL DYNAMICS}

\noindent{\bf OF TRANSPORT BIFURCATIONS.}

\vskip 20pt

\noindent{\bf V.~B. Lebedev and P.~H. Diamond$^\dagger$}

\vskip 20pt

\noindent{\it Physics Department, University of California, 
San Diego, CA92093-0319}

\noindent $^\dagger${\it Also at General Atomics, La Jolla, CA92186-9784}

\vskip 20pt

\noindent{\bf Abstract}

The development and time evolution of a transport barrier in a magnetically confined plasma with non-monotonic, nonlinear dependence of the anomalous flux on mean gradients is analyzed. Upon consideration of both the spatial inhomogeneity and the gradient nonlinearity of the transport coefficient, we find that the transition develops as a bifurcation front with radially propagating discontinuity in local gradient. The spatial location of the transport barrier as a function of input flux is calculated. The analysis indicates that for powers slightly above threshold, the barrier location 
$x_b(t) \sim \bigl( D_n \, t \, (P-P_c)/P_c \bigr)^{1/2},$ 
where $P_c$ is the local transition power threshold and $D_n$ is the neoclassical diffusivity . This result suggests a simple explanation of the high disruptivity observed in reversed shear plasmas. The basic conclusions of this theory are insensitive to the details of the local transport model.   

\vfill
\eject

\noindent {\bf I. Introduction}

\par
Transport barrier physics is a central topic in ongoing research in magnetic fusion. By transport barrier, we refer to a region where anomalous transport is eliminated or very significantly reduced, so confinement is determined by neoclassical processes and macroscopic stability limits. It is important to note that a transport barrier need not  necessarily be a thin layer (such the edge pedestal region in a standard High (H) mode plasma), but rather can extend over a significant fraction of the plasma cross-section (as in the case of Enhanced Reversed Shear (ERS) modes). As intimated above, the spectacular enhanced confinement characteristic of plasmas with transport barriers naturally presents a challenge to stability limits and the particle and exhaust control systems envisioned for advanced tokamak reactors. At the same time, the improved confinement is, of course, highly desirable. Hence, \underbar{control} of transport barriers (and transport in general) is emerging as an!
 important theme for present and
 future research in fusion plasma physics. One example of successful transport barrier control is the Ion Bernstein Wave (IBW) driven Core High (CH) mode achieved on the Modified Princeton Beta Experiment (PBX-M) tokamak. However, it seems fair to say that the science of transport barrier control is still in its infancy.
\par
Understanding the spatio-temporal dynamics of transport barrier is a necessary prerequisite for engineering successful control techniques and algorithms. The many specific questions concerning the space-time characteristics of barrier evolution include:

\item{i.)} \ how can one predict the radial extent and location of a barrier?
  
\item{ii.)} \ what physics enters the criteria for barrier stationarity?

\item{iii.)} \ what is the rate of propagation of a non-stationary transport barrier?

\item{iv.)} \ what are the threshold conditions for barrier formation and relaxation? How much hysteresis is exhibited?

\item{v.)} \ how does a barrier respond to localized secondary external drive? For example, can localized deposition of IBW power be utilized to broaden an ERS transport barrier, thus controlling profile peaking and reducing disruptivity?

\noindent
Here, we pursue several of these questions in the context of a simple, one-field model. 
\par
Many physical problems$^{1}$ require the solution of the non-linear diffusion equation, where the diffusion coefficient $D$ depends on the concentration of the diffused substance $n$. The power-low type of dependence on concentration:
$$
D \sim n^\alpha
$$
leads to the formation of a propagating concentration front with the following behavior:
$$
n(x) \sim \cases{
 \lambda (t)^{-1} \bigl(1-(x/\lambda(t))^2 \bigr)^{1/\alpha},
  &if $|x| \leq \lambda (t)$;\cr
0, &if $|x| > \lambda (t)$,\cr }
$$
(see Fig.1). Here, $\lambda(t) = const \cdot t^{1/(2+\alpha)}.$ In this paper we consider a different type of nonlinear diffusion, namely one for which the diffusivity is a function of the gradient of concentration. Such a functional dependence occurs in magnetic plasma confinement devices, where the anomalous fluxes of high temperature plasma are driven by plasma microinstabilities with the characteristic length scale 
$\lambda \sim \rho_i \ll L_r$. Here $\rho_i$ is the ion Larmour radius and $L_r$ is the typical radial length scale of temperature (or density). In general, the fluxes of heat, particles and momentum are related to the temperature, density and velocity gradients, $\nabla T$, $\nabla n$ and $\nabla V$, through the coefficients of the transport matrix.  In the case of microturbulence driven transport, these coefficients are defined by the amplitude and correlation properties of turbulent fluctuations. The turbulence itself, however, is driven by the microinstabilities associated with the gradients of temperature and density, so the transport matrix is a function of $\nabla T,$ $\nabla n$ and $\nabla V$. Hence, the particle, thermal and momentum fluxes acquire a \underbar{non-linear} dependence on these gradients.

\par
The discovery of improved regimes of plasma confinement, such as H-mode, Very High (VH) mode and the improved core confinement modes$^{2,3,4}$, strongly suggests, that this non-linear dependence is non-monotonic and the fluxes can actually \underbar{decrease} when the density or temperature gradients lie between certain values ({\it i.e.} negative differential diffusivity). As a result, when the rate of fueling of plasma by particles or heat exceeds threshold value, a bifurcation to a new transport regime with higher values of $\nabla n$ or $\nabla T$ takes place$^{5,6,7,8}$. The most plausible physical explanation of this phenomenon is based on the idea of the development of strong radial electric field shear, which stabilizes plasma instabilities and thus decreases the transport during the transition. In equilibrium, the shear of the radial electric field $E_r$ has contributions from the hydrodynamic plasma velocity $V$ and pressure gradient 
$\partial P / \partial r$:
$$
{{ \partial E_r }\over{ \partial r }} = 
- { 1 \over c} { \partial \over{ \partial r }} [{\bf V \times B}]_r + 
{ \partial \over{ \partial r }} \Bigl( { 1 \over {e n}} 
{{ \partial P_i }\over{ \partial r }} \Bigr),
$$
where $P_i$ is the ion pressure and $B$ is the magnetic field. The dynamics of perpendicular plasma velocity ${\bf V}$ are rather complicated$^{8,9,10}$ and thus are beyond the scope of this minimalist paper. Taking bulk velocity to be fixed, the force balance equation shown indicates that the changes in temperature and density gradients will be accompanied by changes in the shear of $E_r$ which, in turn, will change the anomalous transport. This constitutes a minimal model of a transport bifurcation.

\par 
The purpose of this work is to describe the spatial and temporal development of transport bifurcation in the framework of a very simple, general one-field model by studing the geometry of flux surfaces in 
($x, \partial_x n, \Gamma$) space. The model involves the nonlinear diffusion equation with non-monotonic dependence of the local flux on the value of gradient. The model presumes a local transport processes yielding multiple gradients for a certain range of fluxes, \underbar{only}. In particular, no assumptions concerning the transport model or the bifurcation mechanism are involved. We construct the solution in the form of propagating transport bifurcation front, which allows us to determine the final position of  transport barrier from the condition of zero front velocity (\rm{i.e.} stationarity).  
\par
Although this work is discussed in the context of nonlinear transport phenomena in fusion devices, its results can also be relevant to other physical problems in which the diffusive fluxes depend on the gradients non-monotonically. Spinodal decomposition in alloys is one example of such a physical process$^{11}$. Another one is turbulent mass and heat transport in planetary atmospheres with zonal flows. In particular, analysis of the numerical simulations of Jovian atmosphere$^{12}$ show a suppression of radial thermal transport and generation of strong poloidal sheared flows when the radial drop in temperature ({\it i.e.} Rayleigh number) exceeds a certain critical value. This process is a classic example of a transport bifurcation, and is one which very likely determines the spatio-temporal patterns observed in the Jovian atmosphere. 
\par
The dynamics of propagating transport barriers was already discussed in Ref.[13], where the spatially propagating front solutions describing the second order transition from an \underbar{unstable} transport regime to a \underbar{stable} one were introduced. The propagating bifurcation of our present model is different from that. It shares certain common features with front solutions in both the diffusion equation with the nonlinear diffusivity ($D \sim n^\alpha $) and the \underbar{bistable}  reaction-diffusion equation ({\it i.e.} Fitzhugh-Nagumo model)$^{13,14}$. The latter describes the dynamics of the first order transition from a metastble equilibrium state (or phase) to a stable one, corresponding to a global minimum of the Lyapunov function (effective potential energy) associated with the governing equation. The structure of our model equation is different from that classic example, but also has multiple stable equilibrium solutions in a certain range of input flux valu!
es. As in the case of the reacti
on-diffusion equation, the front solution in our system describes the transition between these equilibrium states (or transport ``phases'').
\par
 The remainder of this paper is organized as follows. In Section II we introduce basic equations and study the development of the transport barrier in the model with spatially dependent nonlinear diffusion. In Section III we study the space-time evolution of the propagating transport bifurcation and determine the stationary position of the transport barrier. In Section IV we discuss results, conclusions and their implications.
\vskip 20pt

\noindent {\bf II. Basic Model}

We consider the spatio-temporal dynamics of the simplest, one-dimensional transport system which exhibits the property of negative differential diffusivity:
$$
\partial_t n + \partial_x \Gamma = Q(x),
\eqno{(1)}
$$
where $x$ ($0<x<a$) corresponds to the radial coordinate in a tokamak and the flux $\Gamma$ is given by the following expression:
$$
\Gamma = \Phi (x, \partial_x n) - \epsilon {\ } L(\partial_x) \ast n.
\eqno{(2)}
$$
\par
The detailed form of the function $\Phi (x, \partial_x n)$ is not required here, it need only have the following very general properties$^{5,15}$:
\item{a.} For any fixed value of $\partial_x n$, $\Phi$ decreases with $x$. The rate of this growth increases for $x \geq x_s$ corresponding to an increased level of anomalous transport at the plasma edge;
\item{b.} For any fixed value of $x$, $\Phi$, is a function of $\partial_x n$ only, and has a characteristic S-curve shape (see Fig.2) {\it i.e.} it increases for small values of $\partial_x n$, goes through a maximum, decreases at intermediate values of $\partial_x n$ (corresponding to a negative differential diffusivity) and, finally, increases again for large values of the density gradient. The first interval of increasing $\Phi$ is one of anomalous transport while the second stage of increase is determined by the collisional (neoclassical)  diffusivity. Thus the slopes of $\Phi$ ({\it i.e.} $\delta \Phi / \delta | \partial_x n |$ ) in the two intervals of increasing flux are different.

\par \noindent 
The contourlines of $\Phi (x,n')$ in  the parametric plane $\{ x, n' \}$ are shown on Fig.3, while the behavior of $\Phi$, as a function of parameter $n'$ only, is shown in Fig.2 for several values of $x$. 
\par 
The linear differential operator $L(\partial_x)$ accounts for smoothing effects, which are of higher order in the $\epsilon \sim (\lambda/L_r)^2$ expansion. Thus $L$ regularizes the small scale behavior of $n$. Here, $\lambda$ is the characteristic mixing length in the problem. It is defined either by the correlation length of the microturbulence or by the poloidal larmor radius (in the case of prevailing neoclassical transport). $L$ is assumed to be an odd polynomial in $\partial_x$, starting with the term 
$d \cdot \partial_x^3$. In the first approximation, though, the detailed structure of $L$ is not crucial for describing front propagation, so long as this operator is dissipative. The source term $Q(x)$ is a localized, even function of $x$, centered around $x=0$. 
\par
Let's analyze the stationary solutions of Eq.(1) with $\epsilon=0$:
$$
\partial_x \bigl( \Phi ( x , \partial_x  n) \bigr) = 0
\eqno{(3)}
$$
satisfying the following boundary conditions:
$$
n |_{x=a}=0, \qquad \Gamma |_{x=0} = \Gamma_0.
\eqno{(3a)}
$$
Eq.(3) can be easily integrated to yield:
$$
\Gamma_0 = \Phi( x , \partial_x n).
\eqno{(4)}
$$
This equation is easily analyzed graphically in the plane of parameters $\{ x, n' \}$, where it defines a contourline $\Phi (x, n') = \Gamma_0 = const$. This contourline can be thought of as a plot of the function $n' =  n'(x,\Gamma_0)$ implicitly defined by 
$\Phi (x,n') = \Gamma_0$. It gives the solution of Eq.(3) with the boundary conditions (3a) in the following form:
$$
n_0(x) = \int_a^x n'(s, \Gamma_0) d s.
\eqno{(5)}
$$
Depending on the value of $\Gamma_0$, the function $\Phi$ with the above-described properties allows the following three types of solution:
\item{a).} The solution $n'_1(x, \Gamma_0)$ for small input flux values, $\Gamma_0 < \Gamma_L,$ is shown in Fig.4a. The parameter $\Gamma_L$ is defined as the local minimum of the S-shaped curve, which shows the dependence of the local flux on the density gradient at $x=0$:
$$
\Gamma_L \equiv 
\min \Phi ( 0 , n' ) \quad {\rm for} \quad n'<0;
$$

\item{b).} The solution $n'_3(x,\Gamma_0)$ for large values of flux, $\Gamma_0 > \Gamma_H$, is shown in Fig.4a. The parameter $\Gamma_H$ is defined as the local maximum of the flux curve at $x=a$:
$$
\Gamma_H \equiv 
\max \Phi ( a , n' )  \quad {\rm for} \quad n'<0.
$$ 
In comparison with the previous case, this solution is characterized by a lower level of transport, which is given by the ratio 
$\Gamma_0 / n'_3(x,\Gamma_0)$;

\item{c).} The case with intermediate values of input flux, $\Gamma_H > \Gamma_0 > \Gamma_L$, is shown in Fig.4b. The equation $\Phi ( x, n') = \Gamma_0$ can be inverted with respect to  the parameter $n'$ as $n' =  n'(x,\Gamma_0).$ This function, however, is multivalued in a certain range of $x$:
$0 \leq x_{cr_1}(\Gamma_0) \leq x \leq x_{cr_2}(\Gamma_0) \leq a$ because of the S-shaped form of the corresponding contourline. As a result, we can identify three branches of $n'(x,\Gamma_0)$, which will be further referred to as $n'_i(x), \ i=1,2,3,$ satisfying 
$0 > n'_1 (x,\Gamma_0) > n'_2 (x,\Gamma_0) > n'_3 (x,\Gamma_0)$.
\par\noindent
In case a) or in case b), the solution of the reduced system (3),(3a) given by Eq.(5) approximates the stationary solution of Eq.(1) (when the source term is absent) with the relative accuracy $\epsilon$ everywhere in $x$ except for narrow boundary layers at $x = 0$ and $x = a$. In case c), however, this solution is not well defined because of the presence of several branches of $n'(x,\Gamma_0)$. Nevertheless, it is possible to build a composite solution $n_c(x,\Gamma_0) = \int_a^x n'_c (s,\Gamma_0) d s$, where the function $n'_c$ consists of two or more smooth parts from the different branches of the multivalued function $n'(x,\Gamma_0)$ which are separated by jump discontinuities. In this solution, the branch with larger value of $|n'_c|$ corresponds to the transport barrier. Such a composite solution is a good approximation of the stationary solution of Eq.(1) everywhere in $x$, except for the narrow boundary layers in the vicinity of the jump discontinuities and at the bou!
ndaries $x \rightarrow 0,a$ (see
 Fig.5). These provide smooth transition between the branches corresponding to different transport regimes. The locations of these boundary layers, as well as the possibility of their propagation as dynamical fronts will be discussed in the next Section. One should observe, that the composite solution can be constructed only out of the first and third branches of the multivalued function $n'(x,\Gamma_0)$. Its second branch $n'_2 (x,\Gamma_0)$ belongs to the area of $\{ x, n' \}$ plane which is characterized by \underbar{negative} differential diffusivity:
$$
\Phi'(x) \equiv 
{{ \partial \Phi  (x, n') }\over{ \partial n' }} 
\Bigr|_{n'=n'_2} < 0.
$$
As a result, it is unstable with respect to small perturbations $\tilde n(x,t) \equiv n ( x, t )- n_2(x, \Gamma_0) $. The linearization of Eq.(1) yields
$$
\partial_t \tilde n = 
\partial_x \Bigl( \Phi'(x) \, \partial_x \tilde n \Bigr) 
+ \epsilon L(\partial_x) \ast \tilde n.
$$
The spatial scale of the perturbation, $l,$ is assumed to be small {\it i.e.} 
$l \ll L_r$, but still larger than 
$\sqrt{\epsilon} \, L_r$. The above equation can be written in the following form:
$$
\partial_t \tilde n \approx \Phi' \, \partial^2_x \tilde n.
$$
For $\Phi' < 0$, it produces unstable solutions with small scales growing fastest, in the absence of regularization by $L$. In particular, for an S-shaped $\Phi(x, \partial_x n)$ the instability related to $\Phi' < 0$ value of $\partial_x n$ is what drives the propagation of the bifurcation. The analogous drive for a super-critical bifurcation front is the instability familiar from the Fisher and time dependent Ginzburg-Landau equations, discussed (in the context of transport barrier dynamics) in Ref.[15]. The width of a transitional boundary layer corresponding to a jump in derivative of the composite solution can be estimated from the fact that this layer matches the $n'_1$ and $n'_3$ branches of $n'(x,\Gamma_0)$ through the values of $\partial_x n$ corresponding to a negative differential diffusivity: $(\partial \Phi / \partial n') |_{n'=\partial_x n}$. This is possible only when the change of the nonlinear flux function $\delta \Phi$ accross this layer is balanced by the d!
issipative operator $\epsilon \,
 L$:
$$
\delta \Phi \sim \Gamma_0 \sim \epsilon \, L(\partial_x) \ast n \sim \epsilon d \, {{ \partial_x n }\over{ \Delta_b^2 }}.
$$
This condition immediately yields the following estimate of the boundary layer width $\Delta_b$:
$$
\Delta_b \sim \sqrt{ \epsilon \,
{{ d \, \partial_x n }\over{ \Gamma_0 }} } \sim \epsilon^{1/2} L_r.
$$
\par
The structure of the stationary solutions of Eq.(1) is such that for continuously changing $\Gamma_0$, transitions between different transport regimes occur in the form of bifurcations. As the value of $\Gamma_0$ is increased above $\Gamma_L$, the solution $n_1(x,\Gamma_0)$ with $\Gamma_0 < \Gamma_L$, continuously evolves into the lower branch $n_1(x,\Gamma_0)$ with $\Gamma_0 > \Gamma_L$. This solution continues to change smoothly with the further increase of $\Gamma_0$ until the flux reaches the threshold value corresponding to the local maximum of the flux curve at $x=0$:
$$
\Gamma_L^{thr} \equiv \max \Phi(0,n') \quad {\rm for} \quad n'<0.
$$
When $\Gamma_0 > \Gamma_L^{thr}$, the function $n_1(x,\Gamma_0)$ doesn't exist for $0 < x < x_{cr_1}(\Gamma_0) < a$, so a new solution should be sought either in the form of a composite solution $n_c (x, \Gamma_0)$ or in the form of the branch $n_3(x, \Gamma_0)$. Hence, for $\Gamma_0 \approx \Gamma_L^{thr}$, a small increase of flux results in a significant jump of $-\partial_x n(x, \Gamma_0)$ to a much higher value in the interval  $0 < x < x_{cr_1}(\Gamma_0)$. This bifurcation can be described as a formation of transport barrier. The width of this barrier will be set by the stationary position of the transitional boundary layer connecting the zones of enhanced and suppressed transport. Apparently, this width can't be less than $x_{cr_1}(\Gamma_0)$ and more than $x_{cr_2}(\Gamma_0).$ In principle, when $x_{cr_2}(\Gamma_0) \ge a$ it can cover the whole range of $x$. The issue of the barrier width will be addressed in the Section III in more detail.
 For the values of flux decreasing from some starting value $\Gamma_0$ exceeding $\Gamma_H$, the profile of $n(x, \Gamma_0)$  exhibits a similar bifurcation of the solution $n_3(x,\Gamma_0)$ at $\Gamma_0 = \Gamma_H^{thr}.$ This quantity is defined as the local minimum of the flux curve at $x=a$:
$$
\Gamma_H^{thr} \equiv \min \Phi(a,n') \quad {\rm for} \quad n'<0.
$$
In this case, a transition to the regime with high transport at $x_{cr_2}(\Gamma_0) < x < a$ takes place.
\par
It is clear from the above that the presence of two locally stable branches of the solution $n(x,\Gamma_0)$ for 
$\Gamma_L < \Gamma_0 < \Gamma_H$ results in hysteresis. For example, when $\Gamma_0$ is increased, the transition from $n_1(x,\Gamma_0)$ to a profile with a lower level of transport occurs at $\Gamma_0 = \Gamma_L^{thr} < \Gamma_H,$ while the solution with the profile $n_3(x,\Gamma_0)$ will bifurcate to a profile with a higher level of transport at $\Gamma_0 = \Gamma_H^{thr} > \Gamma_L$, when the flux is decreased. As a result, when $\Gamma_L^{thr} > \Gamma_H^{thr}$ is satisfied, the profile will return \underbar{to} the mode with high level of transport $n_1(x,\Gamma_0)$ at a value of the flux lower than the one that is required for the bifurcation \underbar{from} $n_1(x,\Gamma_0),$ when the flux is increased. Similar hysteresis behavior is also possible for $\Gamma_L^{thr} < \Gamma_H^{thr}$. In this case, the bifurcation to the profile with a lower level of transport will occur at the flux value $\Gamma_L^{thr}$, when the left branching point crosses the left boundary: $x_{cr!
_1}(\Gamma_0)=0$. For the interm
ediate values of flux $\Gamma_H^{thr} > \Gamma_0 > \Gamma_L^{thr}$, the position of the transitional boundary layer $x_b(\Gamma_0)$ is somewhere between $x_{cr_1}(\Gamma_0)$ and $x_{cr_2}(\Gamma_0)$. When the flux is decreased below $\Gamma_L^{thr}$, the transition to the regime with a high level of transport will occur at the moment when $x_b(\Gamma_0)$ crosses the left boundary: $x_b(\Gamma_0) = 0 > x_{cr_1}(\Gamma_0)$. Apparently, this will happen at the value of flux which is lower than $\Gamma_L^{thr}$. These bifurcation scenarios are illustrated in Fig.6a,b. In a simplified model with two different linear diffusivities $D_{an}$ and $D_{neo}$ for $|n'|<|n'_{crit}|$ and $|n'|>|n'_{crit}|$ respectively, the hysteresis ratio (the ratio of the input power necessary for the transition to a higher confinement mode to the power at which the transition back occurs) scales as the ratio of the anomalous to neoclassical diffusivities {\it i.e.} $\Gamma_{L \rightarrow H} / \Gamma_{H !
\rightarrow L} \sim 
D_{an} / D_{neo}.$
\vskip 20pt

\noindent{\bf III. Spatial dynamics of transport barrier.}

When analyzing the dynamics of the solutions with transitional boundary layers, we can neglect the spatial dependence of the nonlinear flux function $\Phi$ as long as the transitional layer width remains small {\it i.e.} $\Delta_x \sim \epsilon^{1/2} \, L_r \ll L_r$. Let's consider the following modification of Eqs. (1),(2) :
$$
\partial_t n + \partial_x \Gamma = 0,
\quad {\rm where} \quad
\Gamma \equiv \Phi(\partial_x n)  - \epsilon d \, \partial_x^3 n.
\eqno{(6)}
$$
The function $\Phi (n')$, which is shown in Fig.7, depends on the variable $n'$ in the same way as does the function $\Phi(x,n')$ with a fixed value of $x$. For function $\Phi$ independent of $x$, the threshold values of $\Gamma_0$ defined in Section II satisfy the following relations:
$$
\Gamma_L^{thr} = \Gamma_H, \quad {\rm and} \quad
\Gamma_H^{thr} = \Gamma_L.
$$
The coefficient $d$ in the last term of the Right Hand Side (RHS) of Eq.(6) is of the order of $(L_r^2 \, \Phi) / \partial_x n$. Equation (6) has the following boundary conditions:
$$
\Phi(\partial_x n|_0) = \Gamma_0, \qquad n|_a = 0,
\qquad \partial_x^3 n|_{\{ 0,a \}} = 0.
\eqno{(7)}
$$
We are interested in the values of flux which allow for multiple stationary solutions of Eq.(6): $\Gamma_H > \Gamma_0 > \Gamma_L$. For these values of $\Gamma_0$, there are two stable, stationary solutions:
$$
n_1(x,\Gamma_0) = n'_1(\Gamma_0) \, (x-a), \quad 
{\rm and} \quad n_3(x,\Gamma_0) = n'_3(\Gamma_0) \, (x-a),
$$
where $n'_{1,3} (\Gamma_0)$ are the first and the third roots of the equation $\Phi(n') = \Gamma_0$ (see Fig.7). If one introduces the parameters $n'_{L,H}$ corresponding to the positions of the local minimum and maximum of the function $\Phi(n')$: $\Phi ( n'_{L,H} ) = \Gamma_{L,H}$, then the inequality 
$-n'_3 ( \Gamma_0 ) > -n'_L > -n'_H > -n'_1 ( \Gamma_0 ) > 0$ is valid.
We look for an asymptotic solution of Eq.(6) with the boundary conditions (7) corresponding to the limit $\epsilon \rightarrow 0.$ This solution should consist of two pieces separated by a boundary layer at the point $x_b$. Each of these pieces lies on a separate, stable branch of the $S$-shaped flux curve $\Phi(n')$:
$$
\cases{
- \partial_x n (x,t) > - n'_L, & for $0<x<x_b-\Delta,$ \cr
- \partial_x n (x,t) < - n'_H, & for $x_b+\Delta<x<a.$ \cr}
$$
Here, $\Delta$ is a free parameter which is small compared to $L_r,$ but exceeds the width of the layer: $\Delta > \Delta_b$. 
Everywhere outside of the transitional layer:  
$x_b-\Delta < x < x_b+\Delta$, the solution is assumed to be a slowly varying function of $x$ with the characteristic length scale $L_r,$ comparable to the system size $a$. 
\par
In a stationary state, the necessary condition for the existence of this solution can be obtained from the integration of the expression for the flux conservation 
$
\Gamma_0 = \Phi( \partial_x n) - \epsilon d \, \partial_x^3 n
$
multiplied by $\partial^2_x n$. The integration yields:
$$
- \Gamma_0 \, \partial_x n \Bigl|_0^a = 
\int_{\partial_x n |_0}^{\partial_x n|_a} 
\Phi(n') \, d n' + {\epsilon \over 2} (\partial_{x}^2 n)^2 \Bigl|_0^a.
\eqno{(8)}
$$
When $\epsilon \rightarrow 0$, the above relation has the following geometrical interpretation: 
the total area between the graphs of $\Gamma \equiv \Gamma_0$ and $\Gamma = \Phi(n')$ over the interval between the points of  their intersection 
$\bigl( n'_1(\Gamma_0), n'_3(\Gamma_0) \bigr)$ is equal to zero. 
(see Fig.7). For every curve $\Phi(n'),$ this relation specifies a single value of $\Gamma_0$. It will be hereafter referred to as $\Gamma_M$. The condition yielding $\Gamma_M$ is, of course, closely related to the Maxwell construction criterion for phase equilibrium.
\par
For $\Gamma_0 \neq \Gamma_M$, a stationary composite profile \underbar{does} \underbar{not} \underbar{exist}. Nevertheless, in this case a \underbar{time dependent} composite profile \underbar{can} be found. It consists of two parts, one with $- \partial_x n < -n'_H$ and the second $- \partial_x n>-n'_L$,  separated by a propagating boundary layer at 
$x \approx x_b(t)$ (see Fig.8). In order to construct such a profile, let's consider a model nonlinear flux function:

$$
\Phi(n') = \cases{ d_a \, n', & - anomalous transport for $- n' \leq -n'_H - \delta$;\cr
d_n \, n', & - neoclassical transport for $- n' \geq -n'_L + \delta$;\cr
g(n'), & -suppressed anomalous transport for $-n'_L - \delta < - n' < -n'_L + \delta$.\cr }
\eqno{(9)}
$$
In this very simple model of nonlinear flux behavior in a tokamak plasmas, the linear diffusion coefficients $d_n$ and $d_a$ correspond to the low level, neoclassical transport in the core ($0 < x < x_b$) and anomalous transport at the edge ($x_b < x < a$). The function $g(n')$ smoothly connects the linear branches of  $\Phi(n')$. Its local maximum and minimum are at the points $n'_H$ and $n'_L,$ correspondingly. In addition, the inequality 
$\Phi(n'_H+\delta) > \Gamma_M > \Phi(n'_L-\delta)$ is assumed to be satisfied, where $\Gamma_M$ is the value of flux obtained from the Maxwell construction for $\Phi(n')$. For small $\epsilon$, the solution with a transitional layer   has the following form:

$$
\partial_t n = 
\cases{
d_n \, \partial^2_x n, \quad {\rm and} \quad 
- \partial_x n < -n'_L - \delta < 0, & 
for $0 < x < x_b(t) - \Delta;$ \cr
d_a \, \partial^2_x n, \quad {\rm and} \quad
0 > - \partial_x n > -n'_H+\delta, & 
for $a > x > x_b(t) + \Delta,$ \cr}
\eqno{(10)}
$$

\noindent
with the following boundary conditions:
$$
\Phi(\partial_x n) \bigl|_0 =\Gamma_0, \qquad
n \bigl|_a =0.
$$
The first two matching conditions at the boundary layer are rather straightforward {\it i.e.} the continuity of $n$:
$$
n \bigl|_{x_b-\Delta} = n \bigl|_{x_b+\Delta} + O(\Delta),
\eqno{(11)}
$$
and the continuity of flux:
$$
\Gamma \bigl|_{x_b-\Delta} = d_n \, \partial_x n \bigl|_{x_b-\Delta} = 
\Gamma \bigl|_{x_b+\Delta} = d_a \, \partial_x n \bigl|_{x_b+\Delta} + 
O( \Delta ).
\eqno{(12)}
$$
An additional matching condition is obtained by integrating Eq.(6) across the transitional layer with the weight $\partial_x n,$
$$
\Gamma \, \partial_x n \Bigl|_{x_b-\Delta}^{x_b+\Delta} = 
\int_{\partial_x n|_{x_b-\Delta}}^{\partial_x n|_{x_b+\Delta}}
\Phi(n') \, dn' + O(\Delta).
\eqno{(13)}
$$
As noted above, this relation is an example of a Maxwell construction$^{16,14,18}$, and is related to the condition for the coexistence of two transport regimes ("phases"). Its geometrical interpretation is identical to that of Eq.(8), shown in Fig.7.
\par
As $\Delta, \epsilon \rightarrow 0$, the above system results in the following linear problem with the surface of discontinuity at $x=x_b(t)$:
$$
\cases{
\partial_t n = d_n \, \partial^2_x n, & for $0< x < x_b(t),$ \cr
\partial_t n = d_a \, \partial^2_x n, & for $a > x > x_b(t),$ \cr}
\eqno{(14)}
$$
$$
n|_a = 0, \qquad \Phi(\partial_x n|_0) = \Gamma_0,
$$
$$
n|_{x_b-0} = n|_{x_b+0}, \qquad 
d_n \, \partial_x n|_{x_b-0} = d_a \, \partial_x n|_{x_b+0} 
\equiv - \Gamma_{x_b},
\eqno{(15)}
$$
$$
\Gamma_{x_b} \, \partial_x n \Bigl|_{x_b-0}^{x_b+0} = 
\int_{\partial_x n|_{x_b-0}}^{\partial_x n|_{x_b+0}}
\Phi(n') \, dn'.
$$
Note, that flux continuity in combination with 
the last condition of transport "phase" equilibrium are equivalent to the definition of the quantity $\Gamma_M$, so the flux at the transitional layer is fixed at 
$ \Gamma_{x_b} \equiv \Gamma_M .$ As a result, these two matching conditions can be rewritten in the following form:
$$
- d_n \, \partial_x n|_{x_b-0} = \Gamma_M, \qquad
- d_n \, \partial_x n|_{x_b+0} = \Gamma_M.
\eqno{(15a)}
$$
\par
When the flux $\Gamma_0$ on the left boundary coincides with $\Gamma_M$, the system (14),(15) gives a trivial solution for $x_b$:
$$
x_b = const, \qquad 0 < x_b < a.
$$
Let's now consider the situation when $\Gamma_0$ slightly exceeds $\Gamma_M$. In that case, $x_b(t)$ can't be constant. Otherwise, the asymptotic (in time) solution of Eq.(14) for $x > x_b$ would be: 
$n(x)=-(\Gamma_M / d_a) \, (x-a),$ giving the value of $n$ at the transitional layer $n(x_b)=(\Gamma_M / d_a) \, (a - x_b) = const.$ However, this value cannot be matched with the $t \rightarrow \infty$ asymptotic of the solution for $0 < x < x_b,$ which increases with time as $n \approx (\Gamma_0 - \Gamma_M) \, t / x_b$. In principle, the system (14),(15) can be solved exactly. Here, we seek an approximate solution, which describes a slow propagation of the transitional layer, {\it i.e.} which satisfies 
$$
\partial_t x_b^2 \ll d_n.
$$
In practice, this is equivalent to requiring that the barrier propagation velocity $v_b = \partial_t x_b$ satisfy $v_b < d_n/x_b.$ The diffusivity $d_a$ exceeds $d_n$ so the relaxation time $\tau_a \sim a^2 / d_a$ for the solution at the edge ($x > x_b(t)$) is small compared to $\tau_n \sim a^2 / d_n$ for the solution in the core ($x < x_b(t)$) to develop. Hence, the solution for $x > x_b(t)$ can be taken to be stationary, {\it i.e.}
$$
n(x) = (\Gamma_M / d_a ) \, (a-x), \quad {\rm for } \ x > x_b(t).
\eqno{(16)}
$$
For $x < x_b(t)$, we can make the following substitution:
$$
n(x) = - { \Gamma_0 \over d_n } \, x +
{{ (\Gamma_0 - \Gamma_M) }\over d_n } \,
{ x^2 \over{ 2 x_b(t) }} + f(x,t),
\eqno{(17)}
$$
where $f$ is a new unknown function satisfying:
$$
\partial_t f = d_n \, \partial^2_x f + 
{{ ( \Gamma_0 - \Gamma_M) }\over{ x_b (t) }} + 
{{ ( \Gamma_0 - \Gamma_M) }\over d_n } 
{ x^2 \over 2 } {{ \partial_t x_b (t) }\over x_b^2},
\eqno{(18)}
$$
$$
\partial_x f|_0 = \partial_x f|_{x_b} = 0.
$$
Here we use the boundary conditions defining the flux at the left boundary, $x=0,$ and at the transitional layer $x=x_b,$ only. According to the assumption of the slow front propagation, the last term in the RHS of Eq.(18) can be neglected. As a result, we obtain the following approximate solution for $x < x_b(t)$:
$$
n(x,t) \approx - {\Gamma_0 \over d_n} \, x \, + 
{{ (\Gamma_0 - \Gamma_M) }\over d_n } \,
{ x^2 \over{ 2 x_b }} + 
(\Gamma_0 - \Gamma_M) \,
\int {{ d t }\over{ x_b(t) }}.
\eqno{(19)}
$$
Matching of this solution with the solution for $x > x_b$ yields the equation for $x_b(t)$:
$$
- {{ \Gamma_0 + \Gamma_M }\over{ 2 d_N }} \, x_b(t) + 
(\Gamma_0 - \Gamma_M) \, \int {{ d t }\over{ x_b (t) }} = 
{ \Gamma_M \over d_a } \, \bigl( a - x_b (t) \bigr).
\eqno{(20)}
$$
This can be easily solved, giving:
$$
x_b(t) = \sqrt{
{{ \Gamma_0 - \Gamma_M }\over{ 2 \Gamma_M }} \,
{{ d_a \, d_n }\over{ d_a - d_n }} \,
(t + C) },
\eqno{(21)}
$$
where $C$ is a constant of integration. The assumption of slow front propagation is satisfied for values of flux $\Gamma_0$ close enough to $\Gamma_M$: 
$(\Gamma_0 - \Gamma_M) / \Gamma_M \ll 1.$ The profile $n(x,t)$, corresponding to the above solution is shown in Fig.9. This simple analysis allows us to conclude, that when the transport barrier is created in a certain range of $x$, it will spread as long as the condition $\Gamma_0 > \Gamma_M$ is satisfied. 
{ \it For $d_a \gg d_n$, the rate of its propagation will be mainly defined by the low diffusion rate in the core:
$ x_b(t) \approx \sqrt{ \bigl( ( \Gamma_0 - \Gamma_M )/(2 \Gamma_M) \bigr) \, d_n t }.$ }
When the $x$-dependence of the nonlinear flux function is taken into account, our analysis may be considered as quasi-local. The conclusions should not significantly change as long as the transition layer width remains smaller than $L_r.$ In the opposite limit of $\partial_t x_b^2 \gg d_n,$ no front solution exists.
\par
The description of the bifurcation scenarios started in Section II can be completed now. For example, when 
$\Gamma_H^{thr} > \Gamma_0 > \Gamma_L^{thr}$, the solution has multiple branches at  
 $0 < x_{cr_1} (\Gamma_0) < x < x_{cr_2} (\Gamma_0) < a,$ the ambiguity in the radial extent of the low transport zone can be resolved. Its  boundary is at the point $x$,
 where the local front velocity $\partial_t x_b,$ obtained from $\Gamma_0$ and the local value of the equilibrium flux $\Gamma_M(x),$ is zero. This takes place, when the local "phase" equilibrium condition is satisfied:
 $\Gamma_0  = \Gamma_M (x).$ Apparently, $\Gamma_M (x_{cr_1}) < \Gamma_0$ and $\Gamma_M (x_{cr_2}) > \Gamma_0$, so such a point can be found in the interval $\bigl( x_{cr_2}, x_{cr_2} \bigr).$
\vskip 20pt

\noindent{\bf IV. Summary and Conclusions.}

Experimental results and theoretical models of the anomalous heat and particle transport in magnetically confined plasmas suggest that they possess the important property of negative differential diffusivity for a range of temperature and density gradients. As a result, when the fueling rate of the plasma exceeds a certain threshold, the profiles undergo a bifurcation to a confinement regime with much steeper gradients ({\it i.e.} transport barrier forms). In this paper we have demonstrated a simple dynamical model of such barrier dynamics. The main conclusions of this paper are summarized below.

\par
i). the spatial localization of the transport barrier is determined by the structure of the nonlinear flux function in $\{ x, \partial_x n \}$ parameter plane. The shape of this function allows for multiple solutions for the profile $n(x)$. The resulting profile is represented by a single solution branch or by a combination of branches corresponding to neoclassical and anomalous transport connected through transition layers. The stationary position $x_b$ of these layers is given by an argument similar to that used in the Maxwell construction. Specifically, the layer is stationary at the point where the total flux through the system coincides with the local value of the Maxwell flux:
$$
\Gamma_0 = \Gamma_M(x).
$$
The flux $\Gamma_M(x)$ is given by the following equation:
$$
\Gamma_M(x) \, \partial_x n \Bigl|_{n'_1}^{n'_3} = 
\int_{n'_1}^{n'_3} \Phi(n',x) \, dn',
\quad {\rm where} \quad
\Phi(n'_{ \{ 1,3 \} },x) = \Gamma_M (x).
$$

\par
ii). the spatial dynamics of the transition evolves in the form of a propagating transition layers. For $\partial_t x_b^2 < d_n$, the speed of their propagation is defined 
 by the neoclassical diffusivity and the difference between the input flux $\Gamma_0$ and the local Maxwell flux:
$$
x_b(t) \sim \sqrt{ {{ \Gamma_0 - \Gamma_M }\over \Gamma_M } \,
d_n \, t }. 
$$
It seems likely, that  for values of flux $\Gamma_0$ significantly different from $\Gamma_M$, $|\Gamma_0-\Gamma_M| \sim \Gamma_M,$ the front velocity is still restricted by the slowest (neoclassical) diffusion  rate. This follows from the observation, that for $0<x<x_b$ profile evolves on a time scale $x^2_b/d_n,$ which should be comparable with the time scale of the front propagation, {\it i.e.} $x_b/ \partial_t x_b.$

\par
iii). the transition between different transport phases exhibits hysteresis. Even in the simple model considered in our paper, several types of hysteresis are possible. It occurs both  when the flux $\Gamma_0$ is increased beyond  $\Gamma_L^{thr}$ and subsequently reduced, and when $\Gamma_0$ is increased above $\Gamma_H > \Gamma_L^{thr}$ and later reduced. 
\par
The transport barrier propagation phenomenon described here is somewhat similar to the spatial evolution of a first order transition from a metastable phase to a stable phase in a non-equilibrium thermodynamic system. The condensation of overheated vapor$^{17}$ or the spinodal decomposition$^{12}$ are examples of such processes. Our results may also be relevant to the anomalous transport in convective systems, such as planetary atmospheres, where the turbulence coexists with large scale, self-consistently generated zonal flows. 
\par 
As mentioned in the introduction, understanding the space-time response of transport barriers is critical to transport control, which is a likely element of any realization of an advanced tokamak. In this regard, one particularly compelling motivation for transport control is to reduce and minimize the disruptivity of core transport barrier plasmas thus allowing full exploitation of their improved confinement. The disruptions which occur in core barrier plasmas are almost certainly a consequence of the dramatic steepening of the core pressure gradient due to barrier formation at high heating power. In the context of the simple model discussed here and in Ref.[15], the core pressure gradient is determined by:

\par a). the source strength and (fixed) anomalous diffusivity,

\par b). the transition layer location $x_b(t)$, which supplies the "boundary" condition. 

\noindent
In simple terms, our model suggests that the global profile evolution will resemble swinging doors attached to a sliding hinge, located at $x_b(t)$. Thus, the energy content can either be increased by $x_b(t)$ moving outward or by increasing $P'_{core}$, eventually leading to violation of macroscopic stability limits. Hence, the dynamics and stationary value of $x_b(t)$play an important role in determining disruptivity! One of the central results of this analysis is that the hinge cannot slide very fast ({\it i.e.} 
$x_b(t) \sim \bigl( d_n t \, (P -P_c)/P_c \bigr)^{1/2}$). This situation is exacerbated by the radial dependence of $P_c,$ which increases sharply with $r$ as $\tilde s$ goes positive$^{15}.$ Since the barrier cannot expand quick enough, $P'_{core}$ surges due to the combined effects of strong fueling and low transport, which in turn steepens the profile and leads to disruption. A straightforward corollary of this hypothesis  is that disruptivity should decrease for broader barriers. This is borne out in Doublet III-D (DIII-D) Negative Core Shear (NCS) discharges, where an L$\rightarrow$H transition is triggered (thus broadening the deposition profile), and Weak Negative Shear (WNS) plasmas, in which the $q(r)$ profile is rather flat, thus reducing the radial gradient in $P_{core}(r).$ It should be noted, however, that control via an H-mode edge is undesirable, since the beneficial properties of the L-mode are lost. Alternative control schemes, which exploit Radio Frequency (R!
F)-driven shear layers to broade
n the transition barrier are being examined$^{18}$ and will be discussed in future publications. 

\noindent{\bf Acknowledgment.}

We thank B.~A. Carreras, D.~E. Newman and K.~H. Burrell for useful conversations and suggestions. This work is supported by the United States Department of Energy Grants No. DE-FG03-88ER53275 and the Office of Naval Research Grant No. N00014-91-J-1127. 

\vfill\eject

\noindent{\bf References.}

\par\noindent $^1$ M.~N. Ozisik {\it Heat Conduction} 2nd ed., (John Wiley and Sons: New York, 1993)
\par\noindent $^2$ ASDEX Team, Nucl. Fusion {\bf 29}, 1959 (1989)
\par\noindent $^3$ E.~J. Strait, L.~L. Lao, M.~E. Mauel, B.~W. Rice, T.~S. Taylor, 
K.~H. Burrell, M.~S. Chu, E.~A. Lazarus, T.~H. Osborne, S.~J. Thompson, A.~D. Turnbull, Phys. Rev. Lett. {\bf 75}, 4421 (1995)
\par\noindent $^4$ F.~M. Levinton, M.~C. Zarnstorff, S.~H. Batha, M. Bell, R.~E. Bell, R.~V. Budny, C. Bush, Z. Chang, E. Fredrickson, A. Janos, J. Manickam, A. Ramsey, S.~ Sabbagh, G.~L. Schmidt, E.~J. Synakowski, G. Taylor,  Phys. Rev. Lett. {\bf 75}, 4417 (1995)
\par\noindent $^5$ F.~L. Hinton, Phys. Fluids B, {\bf 3}, 696 (1991)
\par\noindent $^6$ B.~A. Carreras, D. Newman, P.~H. Diamond, Y.~M. Liang, Phys. Plasmas {\bf 1}, 4014 (1994)
\par\noindent $^7$ B.~B. Kadomtsev, {\it Tokamak Plasma: A Complex Physical System} (Institute of Physics Publishing, Bristol and Phyladelphia, 1992)
\par\noindent $^8$ P.~H. Diamond, V.~B.  Lebedev, Y.-M. Liang, A.~V. Gruzinov, I. Gruzinova, M. Medvedev, B.~A. Carreras, D.~E. Newman, L. Charlton, K.~L. Sidikman, D.~B. Batchelor, E.~F. Jaeger, C.~Y. Wang, G.~G. Craddock, N. Mattor, T.~S. Hahm, M. Ono, B. LeBlanc, H. Biglari, F.~Y. Gang, D.~J. Sigmar
``Dynamics of the L-H Transition, VH-Mode Evolution, Edge Localized Modes and R.F. Driven Confinement Control in Tokamaks'', in the 
{\it Proc. 15th Int. Conf. on Plasma
Physics and Controlled Nuclear Fusion Research, Seville, Spain} (1994)
\par\noindent $^9$ P.~H. Diamond, Y.~M. Liang, B.~A. Carreras, P.~W. Terry, Phys. Rev. Lett. {\bf 72}, 2565 (1994)
\par\noindent $^{10}$ V.~B. Lebedev, P.~N. Yushmanov, P.~H. Diamond, S.~V. Novakovskii, A.~I. Smolyakov, Phys. Plasmas {\bf 3}, 3023 (1996)
\par\noindent $^{11}$ H. Furukawa, Advances in Physics {\bf 34}, 703 (1996)
\par\noindent $^{12}$ A.~C. Or, F.~H. Busse, J. Fluid. Mech. {\bf 174}, 313 (1987)
\par\noindent $^{13}$ P.~H. Diamond, V.~B. Lebedev, D.~E. Newman, B.~A. Carreras, Phys. Plasmas {\bf 2}, 3685 (1994)
\par\noindent $^{14}$ J.~D. Murray, {\it Mathematical Biology} (Springer-Verlag, Berlin, Heidelberg, 1989)
\par\noindent $^{15}$ P.~H. Diamond, V.~B. Lebedev, D.~E. Newman, B.~A. Carreras, T.~S. Hahm, W.~M Tang, G. Rewoldt, K. Avinash, "On the dynamics of transition to enhanced confinement in reversed magnetic shear discharges", submitted to Phys. Rev. Lett., (1996)
\par\noindent $^{16}$ G.~B. Whitham, {\it Nonlinear Waves} (New York: Academic Press 1974)
\par\noindent $^{17}$ L.~D. Landau, E.~M. Lifshitz, {\it Statistical physics} 2nd ed., (Pergamon Press, London, 1969)
\par\noindent $^{18}$ B. LeBlanc, S. Batha, R. Bell, S. Bernabei, L. Blush, 
E. de la Luna, R. Doerner, J. Dunlap, A. England, I. Garcia, D. Ignat, R. Isler, S. Jones, R. Kaita, S. Kaye, H. Kugel, F. Levinton, S. Luckhardt, T. Mutoh, M. Okabayashi, M. Ono, F. Paoletti, S. Paul, G. Petravich, A. Post-Zwicker, N. Sauthoff, L. Schmitz, S. Sesnic, H. Takahashi, M. Talvard, W. Tighe, G. Tynan, S. von Goeler, P. Woskov, and A. Zolfaghari, Phys. Plasmas {\bf 2}, 741 (1995)
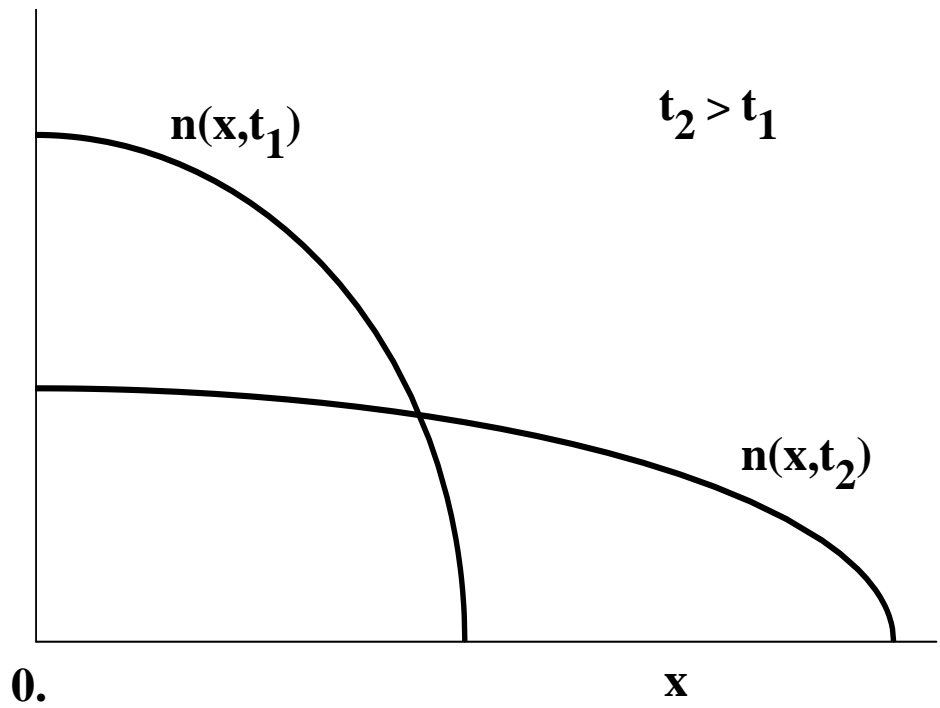
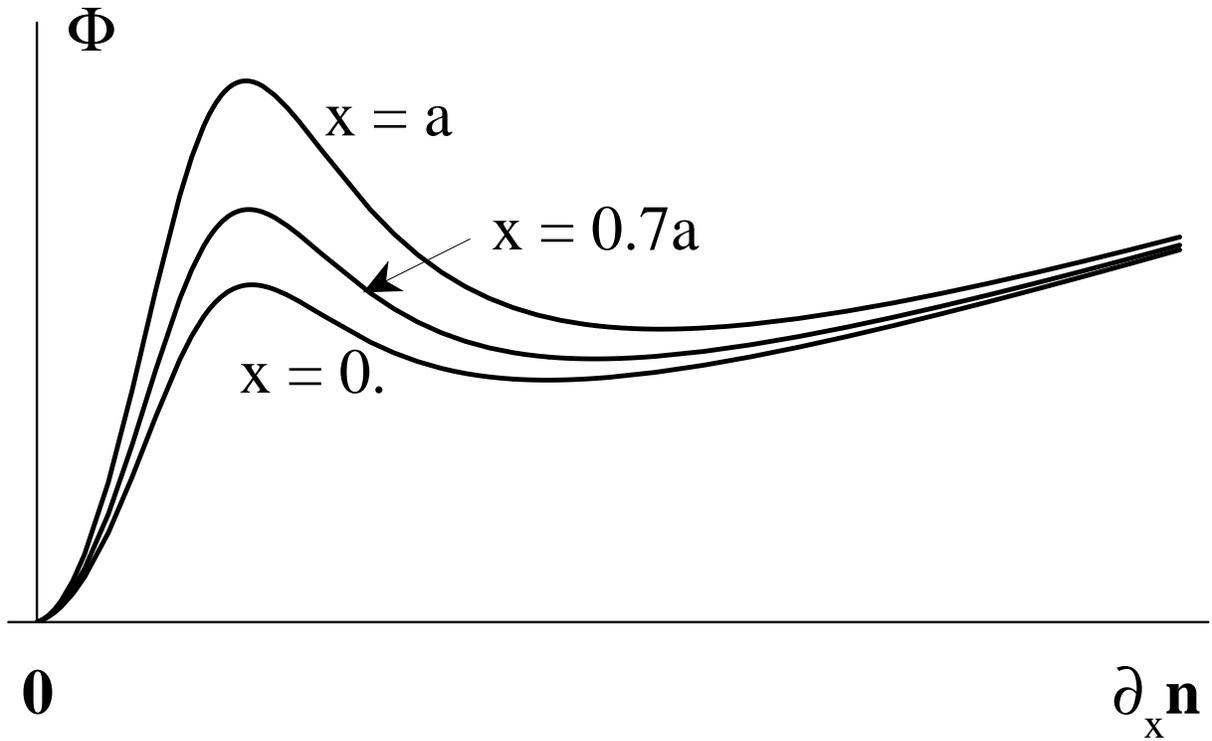
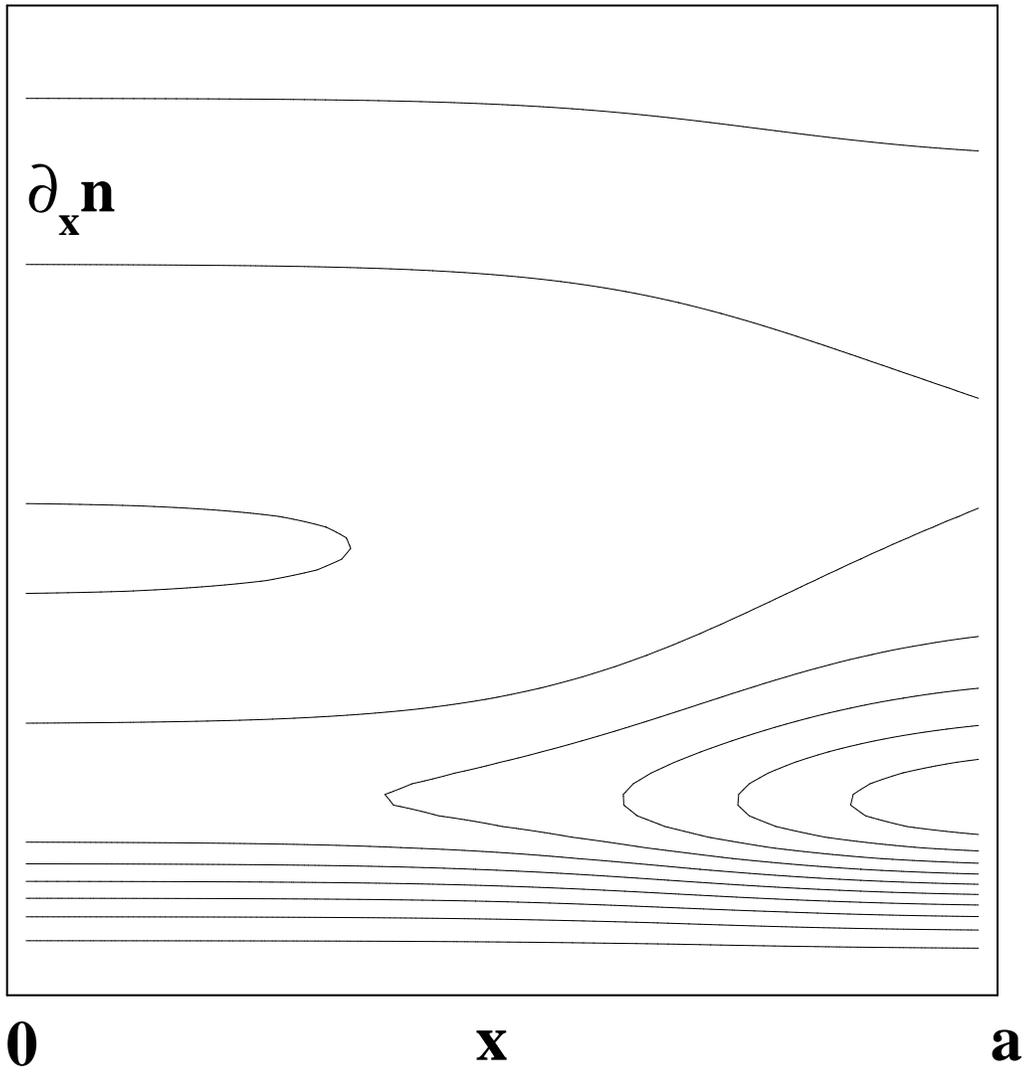
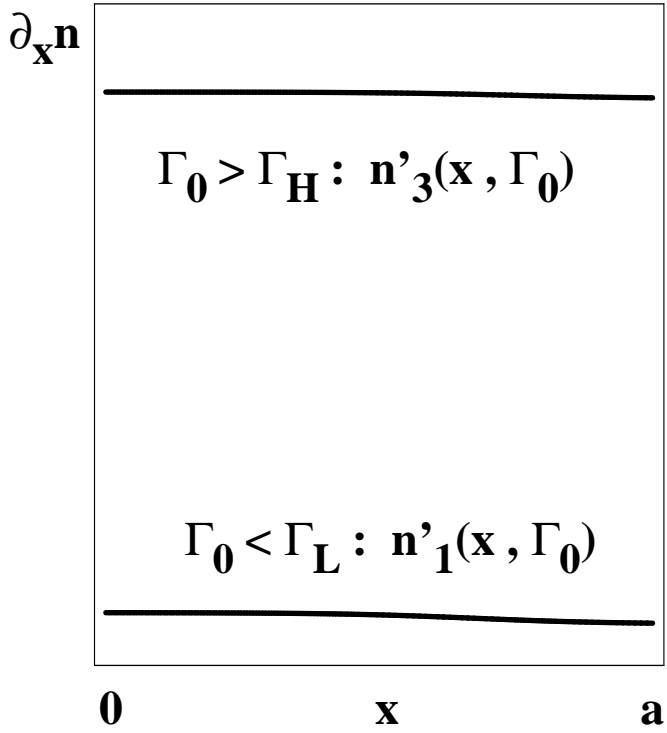
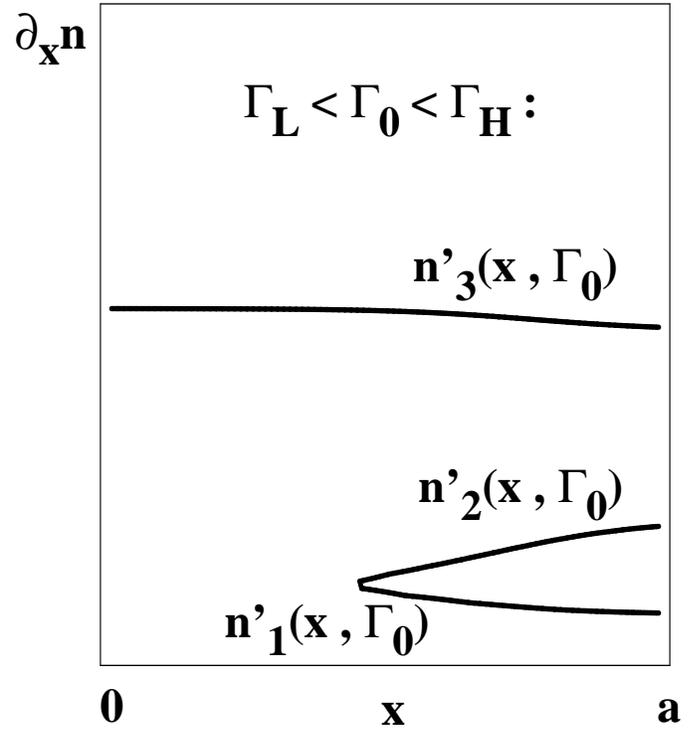
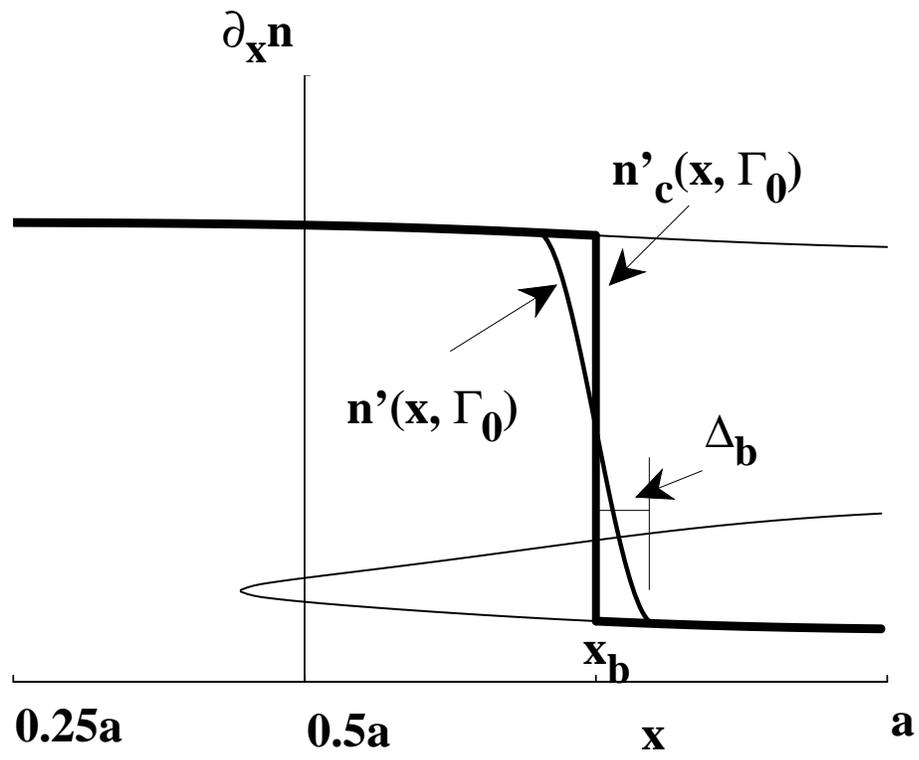
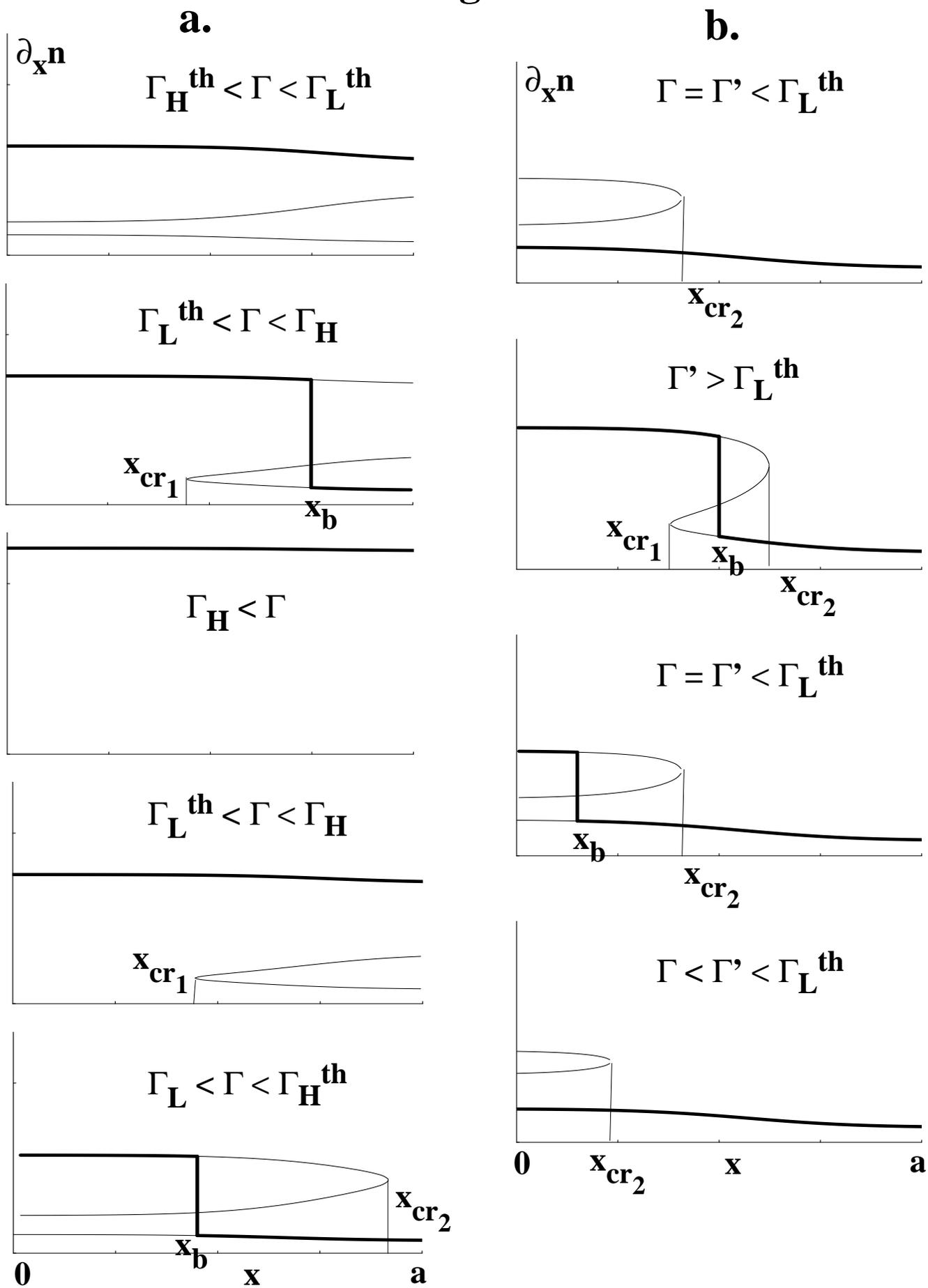
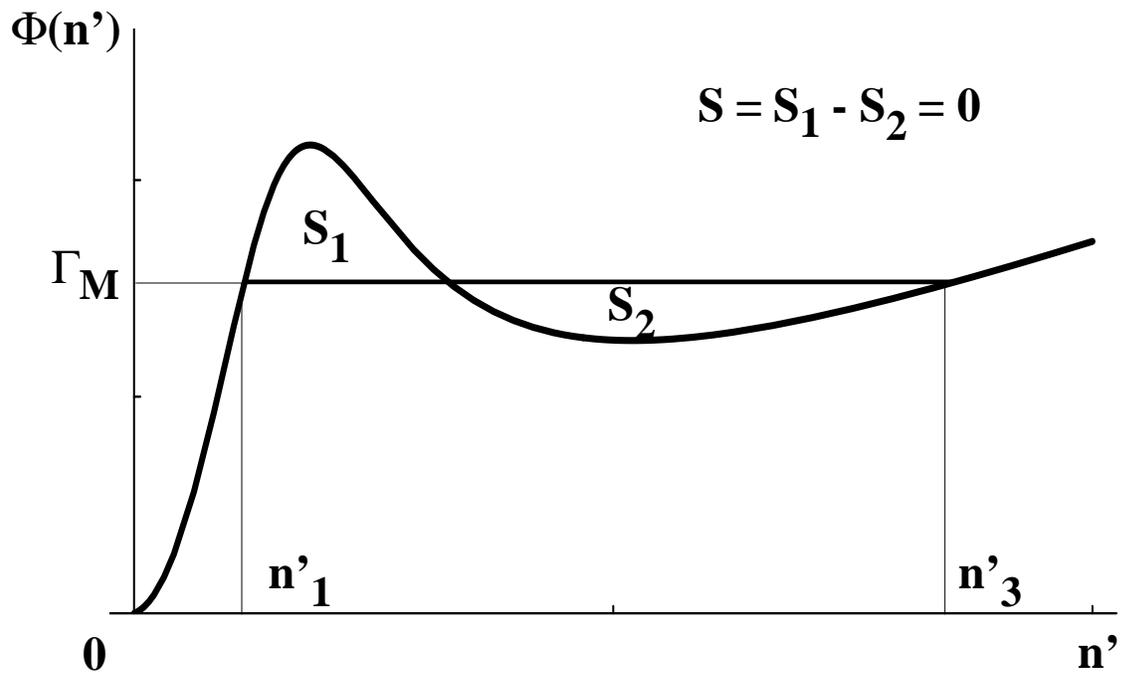
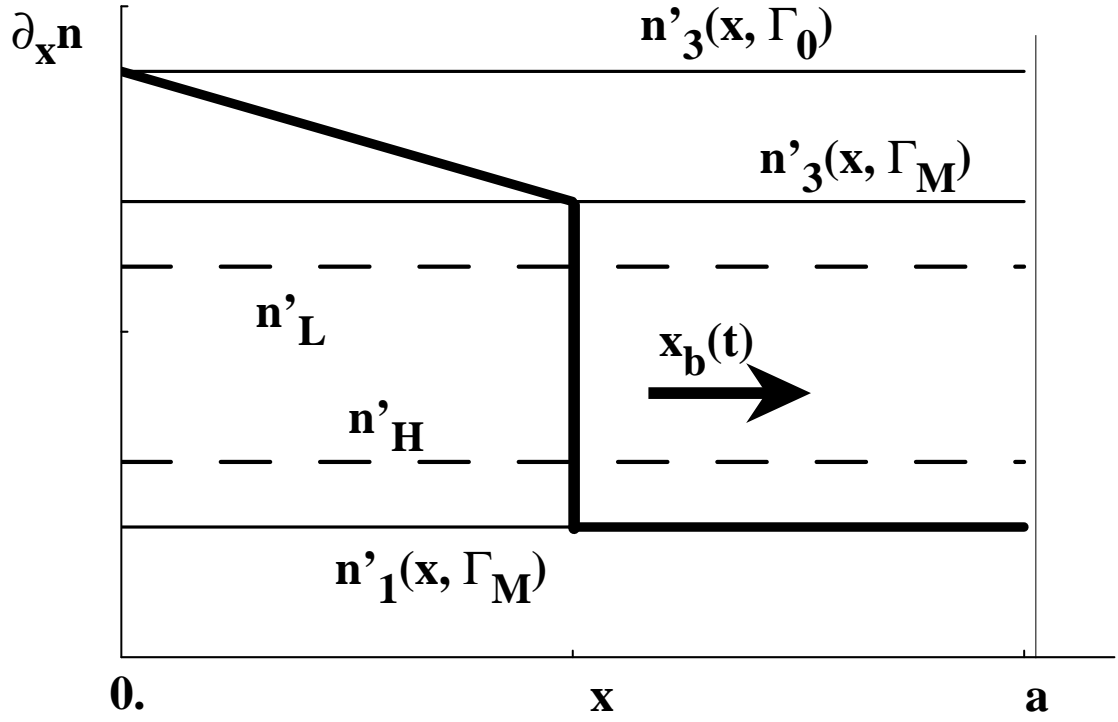
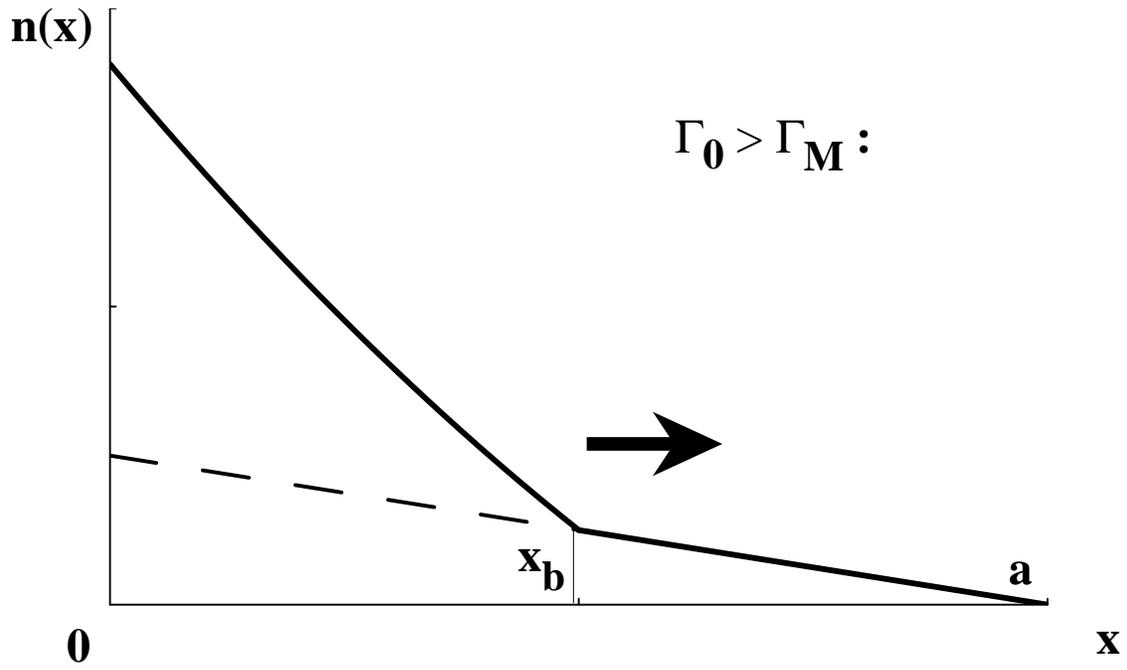
\end